# Complexity of Determining Nonemptiness of the Core[*]

**Vincent Conitzer** and **Tuomas Sandholm**

Carnegie Mellon University

Computer Science Department

5000 Forbes Avenue

Pittsburgh, PA 15213, USA

{conitzer,sandholm}@cs.cmu.edu

## Abstract

Coalition formation is a key problem in automated negotiation among self-interested agents, and other multiagent applications. A coalition of agents can sometimes accomplish things that the individual agents cannot, or can do things more efficiently. However, motivating the agents to abide to a solution requires careful analysis: only some of the solutions are stable in the sense that no group of agents is motivated to break off and form a new coalition. This constraint has been studied extensively in cooperative game theory. However, the computational questions around this constraint have received less attention. When it comes to coalition formation among software agents (that represent real-world parties), these questions become increasingly explicit.

In this paper we define a concise general representation for games in characteristic form that relies on superadditivity, and show that it allows for efficient checking of whether a given outcome is in the core. We then show that determining whether the core is nonempty is $\mathcal{NP}$-complete both with and without transferable utility. We demonstrate that what makes the problem hard in both cases is determining the collaborative possibilities (the set of outcomes possible for the grand coalition), by showing that if these are given, the problem becomes tractable in both cases. However, we then demonstrate that for a hybrid version of the problem, where utility transfer is possible only within the grand coalition, the problem remains $\mathcal{NP}$-complete even when the collaborative possibilities are given.

## 1 Introduction

Coalition formation is a key problem in automated negotiation among self-interested agents. A coalition of agents can sometimes accomplish things that the individual agents cannot, or can do things more efficiently. However, motivating the agents to abide to a solution requires careful analysis: only some of the solutions are *stable* in the sense that no group of agents is motivated to break off and form a new coalition. This constraint has been studied extensively in cooperative game theory. However, the computational questions around this constraint have received less attention. When it comes to coalition formation among software agents (that represent real-world parties), these questions become increasingly explicit.

In general, computational complexity could stem from each potential coalition having some hard optimization problem. For example, when the agents are carrier companies with their own trucks and delivery tasks, they can save costs by forming a coalition (pooling their trucks and tasks), but each potential coalition faces a hard optimization problem: a vehicle routing problem defined by the coalition's trucks and tasks. The effect of such hard optimization problems on coalition formation has been studied by Sandholm and Lesser [Sandholm and Lesser, 1997]. As in the bulk of research on coalition formation, in this paper we do not address that issue. Rather, we assume that such optimization problems have already been solved (at least the pertinent ones), and given this, we characterize the stable feasible outcomes. This has been the focus of most of the work in coalition formation. The contribution of this paper belongs to the relatively new, small set of papers that study the complexity of characterizing such solutions.

The determination of stable solutions has electronic commerce applications beyond automated negotiation as well. For example, consider a large number of companies, some subsets of which could form profitable virtual organizations that can respond to larger or more diverse orders than the individual companies can. Determining stable value divisions allows one to see which potential virtual organizations would be viable in the sense that the companies in the virtual organization would naturally stay together. As another example, consider a future online service that determines how much each employee of a company should be paid so that the company does not collapse as a result of employees being bought away by other companies. The input to this service would be how much subsets of the company's employees would be paid if they left collectively (for instance, a whole department could be bought away). This input could come from salary databases or a manager's estimate. The computational problem of determining a stable renumeration would be crucial for such a service. Both of these example problems fit exactly under the model that we study in this paper.

---

[*]The material in this paper is based upon work supported by the National Science Foundation under CAREER Award IRI-9703122, Grant IIS-9800994, ITR IIS-0081246, and ITR IIS-0121678.

The rest of the paper is organized as follows. In Section 2, we review the required concepts from cooperative game theory. In Section 3, we define a concise general representation for games in characteristic form that relies on superadditivity, and show that it allows for efficient checking of whether a given outcome is in the core. In Section 4, we show that determining whether the core is nonempty is $\mathcal{NP}$-complete both with and without transferable utility. In Section 5, we demonstrate that what makes the problem hard in both cases is determining the collaborative possibilities (the set of outcomes possible for the grand coalition), by showing that if these are given, the problem becomes tractable in both cases. In Section 6, we show that for a hybrid version of the problem, where utility transfer is possible only within the grand coalition, the problem remains $\mathcal{NP}$-complete even when the collaborative possibilities are given.

## 2 Definitions from cooperative game theory

In this section we review standard definitions from cooperative game theory, which we will use throughout the paper. In the definitions, we follow the most prevalent advanced textbook in microeconomics [Mas-Colell *et al.*, 1995].

In general, how well agents in a coalition do may depend on what nonmembers of the coalition do (e.g. [Bernheim *et al.*, 1987; Chatterjee *et al.*, 1993; Evans, 1997; Milgrom and Roberts, 1996; Moreno and Wooders, 1996; Okada, 1996; Ray, 1996]). However, in cooperative game theory, coalition formation is usually studied in the context of *characteristic function games* where the utilities of the coalition members do not depend on the nonmembers' actions [Kahan and Rapoport, 1984; van der Linden and Verbeek, 1985; Zlotkin and Rosenschein, 1994; Charnes and Kortanek, 1966; Shapley, 1967; Wu, 1977]. (One way to interpret this is to consider the coalition members' utilities to be the utilities they can *guarantee* themselves no matter what the nonmembers do [Aumann, 1959; Tohmé and Sandholm, 1999].)

**Definition 1** *Given a set of players $A$, a* utility possibility vector $u^B$ *for* $B = \{b_1, \ldots, b_{n_B}\} \subseteq A$ *is a vector* $(u_{b_1}, \ldots, u_{b_{n_B}})$ *representing utilities that the players in $B$ can guarantee themselves by cooperating with each other. A* utility possibility set *is a set of utility possibility vectors for a given set $B$.*

**Definition 2** *A* game in characteristic form *consists of a set of players $A$ and a utility possibility set $V(B)$ for each $B \subseteq A$.*

Sometimes games in characteristic form have *transferable utility*, which means agents in a coalition can transfer utility among themselves.

**Definition 3** *A game in characteristic form is said to have* transferable utility *if for every $B \subseteq A$ there is a number $v(B)$ (the* value *of $B$) such that*
$V(B) = \{u^B = (u_{b_1}^B, \ldots, u_{b_{n_B}}^B) : \sum_{b \in B} u_b^B \leq v(B)\}.$

It is commonly assumed that the joining of two coalitions does not prevent them from acting as well as they could have acted separately. In other words, the composite coalition can coordinate by choosing not to coordinate. This assumption is known as *superadditivity*.[1] We will assume superadditivity throughout the paper. This actually makes our hardness results *stronger* because even a restricted version of the problem is hard.

**Definition 4** *A game in characteristic form is said to be* superadditive *if, for any $B, C \subseteq A$ with $B$ and $C$ disjoint, and for any $u^B \in V(B)$ and $u^C \in V(C)$, we have $(u^B, u^C) \in V(B \cup C)$. (In the case of transferable utility, this is equivalent to saying that for any $B, C \subseteq A$ with $B$ and $C$ disjoint, $v(B \cup C) \geq v(B) + v(C)$.)*

We now need a solution concept. In this paper, we study only the best known solution concept, which is called the *core* [Mas-Colell *et al.*, 1995; Kahan and Rapoport, 1984; van der Linden and Verbeek, 1985]. It was first introduced by Gillies [Gillies, 1953].

**Definition 5** *An* outcome $u^A = (u_1^A, \ldots, u_n^A) \in V(A)$ *is* blocked *by coalition $B \subseteq A$ if there exists* $u^B = (u_{b_1}^B, \ldots, u_{b_{n_B}}^B) \in V(B)$ *such that for all $b \in B$, $u_b^B > u_b^A$. (In the case of transferable utility, this is equivalent to saying that the outcome is blocked by $B$ if $v(B) > \sum_{b \in B} u_b^A$.) An outcome is in the* core *if it is blocked by no coalition.*

In general, the core can be empty. If the core is empty, the game is inherently unstable because no matter what outcome is chosen, some subset of agents is motivated to pull out and form their own coalition. In other words, requiring that no subset of agents is motivated to break off into a coalition of its own overconstrains the system.

An example of a game with an empty core is the one with players $\{x, y, z\}$, where we have the utility possibility vectors $u^{\{x,y\}} = (2, 1), u^{\{y,z\}} = (2, 1)$, and $u^{\{x,z\}} = (1, 2)$ (and the ones that can be derived from this through superadditivity). The same example with transferable utility also has an empty core.

In the rest of this paper, we will study the question of how complex it is to determine whether the core is nonempty, that is, whether there is a solution or the problem is overconstrained.

## 3 Representing characteristic form games concisely

In our representation of games in characteristic form, we distinguish between games without transferable utility, where we specify some utility possibility vectors for some coalitions, and games with transferable utility, where we specify the values of some coalitions.

---
[1] When superadditivity holds, it is always best for the grand coalition of all agents to form. On the other hand, without superadditivity, even finding the optimal coalition structure (partition of agents into coalitions) can be hard [Sandholm *et al.*, 1999; Larson and Sandholm, 2000; Shehory and Kraus, 1998; Shehory and Kraus, 1996; Ketchpel, 1994].

If the representation of the game specifies $V(B)$ or $v(B)$ explicitly for each coalition $B \subseteq A$, then the length of the representation is exponential in the number of agents. In that case, any algorithm for evaluating nonemptiness of the core (as long as it reads all the input) requires time exponential in the number of agents. However, that run time is polynomial in the size of the input (this can be accomplished, for example, using the algorithms that we introduce in Section 5).

Of course, most characteristic form games that represent real-world settings have some special structure. This usually allows for a game representation that is significantly more concise. The complexity of characterizing the core has already been studied in certain very specific concisely expressible families of games before. For example, Faigle et al. study the complexity of testing membership in the core in minimum cost spanning tree games [Faigle *et al.*, 1994]. Deng and Papadimitriou study games where the players are nodes of a graph with weights on the edges, and the value of a coalition is determined by the total weight of the edges contained in it [Deng and Papadimitriou, 1994]. Deng et al. study an integer programming formulation which captures many games on graphs [Deng *et al.*, 1997]. All of those results depend heavily on concise game representations which are specific to the game families under study. Typically, such a family of games is played on a combinatorial structure. Cooperative games on combinatorial structures have been systematically studied [Bilbao, 2000].

As a point of deviation, we study a natural representation that can capture *any* characteristic form game.[2] Conciseness in our representation stems only from the fact that in many settings, the synergies among coalitions are sparse. When a coalition introduces no new synergy, its utility possibility vectors can be *derived* using superadditivity. Therefore, the input needs to include only the utility possibility vectors of coalitions that introduce synergy. The following definitions make this precise.

**Definition 6** *We represent a game in characteristic form without transferable utility by a set of players $A$, and a set of utility possibility vectors $W = \{(B, u^{B,k})\}$. (Here there may be multiple vectors for the same $B$, distinguished by different $k$ indices.) The utility possibility set for a given $B \subseteq A$ is then given by $V(B) = \{u^B : u^B = (u^{B_1}, \ldots, u^{B_r}), \bigcup_{1 \leq i \leq r} B_i = B$, all the $B_i$ are disjoint, and for all the $B_i$, $(B_i, u^{B_i}) \in W\}$. To avoid senseless cases that have no outcomes, we also require that $(\{a\}, (0)) \in W$ for all $a \in A$.[3]*

**Definition 7** *We represent a game in characteristic form with transferable utility by a set of players $A$, and a set of values $W = \{(B, v(B))\}$. The value for a given $B \subseteq A$ is then given by $v(B) = \max\{\sum_{1 \leq i \leq r} v(B_i) : \bigcup_{1 \leq i \leq r} B_i = B$, all the $B_i$ are disjoint, and for all the $B_i$, $(B_i, v(B_i)) \in W\}$. To avoid senseless cases that have no outcomes, we also require that $(\{a\}, 0) \in W$ whenever $\{a\}$ does not receive a value elsewhere in $W$.*

So, we only need to specify a basis of utility possibilities, from which we can then derive the others. This representation integrates rather nicely with real-world problems where determining any coalition's value is complex. For example, in the multiagent vehicle routing problem, we solve the routing problem for every coalition that might introduce new synergies. When it is clear that there is no synergy between two coalitions (for example, if they operate in different cities and each one only has deliveries within its city), there is no need to solve the routing problem of the coalition that would result if the two coalitions were to merge.

The following lemmas indicate that we can also use this representation effectively for checking whether an outcome is in the core, that is, whether it satisfies the strategic constraints.

**Lemma 1** *Without transferable utility, an outcome $u^A = (u_1^A, \ldots, u_n^A) \in V(A)$ is blocked by some coalition if and only if it is blocked by some coalition $B$ through some utility vector $u^B$, where $(B, u^B) \in W$.*

**Proof**: The "if" part is trivial. For "only if", suppose $u^A$ is blocked by coalition $C$ through some $u^C$, so that for every $c \in C$, $u_c^C > u_c^A$. We know $u^C = (u^{C_1}, \ldots, u^{C_r})$ where $(C_i, u^{C_i}) \in W$. But then, $C_1$ blocks $u^A$ through $u^{C_1}$. ∎

The proof for the same lemma in the case of transferable utility is only slightly more intricate.

**Lemma 2** *With transferable utility, an outcome $u^A = (u_1^A, \ldots, u_n^A) \in V(A)$ is blocked by some coalition if and only if it is blocked by some coalition $B$ through its value $v(B)$, where $(B, v(B)) \in W$.*

**Proof**: The "if" part is trivial. For "only if", suppose $u^A$ is blocked by coalition $C$ through $v(C)$, so that $v(C) > \sum_{c \in C} u_c^A$. We know that $v(C) = \sum_{1 \leq i \leq r} v(C_i)$ where $(C_i, v(C_i)) \in W$. It follows that $\sum_{1 \leq i \leq r} v(C_i) > \sum_{1 \leq i \leq r} \sum_{c \in C_i} u_c^A$, and hence for at least one $C_i$, we have $v(C_i) > \sum_{c \in C_i} u_c^A$. But then $C_i$ blocks $u^A$ through $v(C_i)$. ∎

## 4 Checking whether the core is nonempty is hard

We now show that with this representation, it is hard to check whether the core is nonempty. This holds both for the non-transferable utility setting and for the transferable utility setting.

**Definition 8 (CORE-NONEMPTY)** *We are given a superadditive game in characteristic form (with or without transferable utility) in our representation language. We are asked whether the core is nonempty.*

---

[2]Our hardness results are not implied by the earlier hardness results for specific game families because it is not possible to concisely represent those games in our input language.

[3]Setting the utility to 0 in this case is without loss of generality, as we can simply normalize the utility function to obtain this.

We will demonstrate $\mathcal{NP}$-hardness of this problem by reducing from the $\mathcal{NP}$-complete EXACT-COVER-BY-3-SETS problem [Garey and Johnson, 1979].

**Definition 9 (EXACT-COVER-BY-3-SETS)** *We are given a set $S$ of size $3m$ and a collection of subsets $\{S_i\}_{1 \leq i \leq r}$ of $S$, each of size 3. We are asked if there is a cover of $S$ consisting of $m$ of the subsets.*

We are now ready to state our results.

**Theorem 1** *CORE-NONEMPTY without transferable utility is $\mathcal{NP}$-complete.*

**Proof**: To show that the problem is in $\mathcal{NP}$, nondeterministically choose a subset of $W$, and check if the corresponding coalitions constitute a partition of $A$. If so, check if the outcome corresponding to this partition is blocked by any element of $W$.

To show $\mathcal{NP}$-hardness, we reduce an arbitrary EXACT-COVER-BY-3-SETS instance to the following CORE-NONEMPTY instance. Let the set of players be $A = S \cup \{w, x, y, z\}$. For each $S_i$, let $(S_i, u^{S_i})$ be an element of $W$, with $u^{S_i} = (2, 2, 2)$. Also, for each $s \in S$, let $(\{s, w\}, u^{\{s,w\}})$ be an element of $W$, with $u^{\{s,w\}} = (1, 4)$. Also, let $(\{w, x, y, z\}, u^{\{w,x,y,z\}})$ be an element of $W$, with $u^{\{w,x,y,z\}} = (3, 3, 3, 3)$. Finally, let $(\{x, y\}, u^{\{x,y\}})$ with $u^{\{x,y\}} = (2, 1)$, $(\{y, z\}, u^{\{y,z\}})$ with $u^{\{y,z\}} = (2, 1)$, $(\{x, z\}, u^{\{x,z\}})$ with $u^{\{x,z\}} = (1, 2)$ be elements of $W$. The only other elements of $W$ are the required ones giving utility 0 to singleton coalitions. We claim the two instances are equivalent.

First suppose there is an exact cover by 3-sets consisting of $S_{c_1}, \ldots, S_{c_m}$. Then the following outcome is possible: $(u^{S_{c_1}}, \ldots, u^{S_{c_m}}, u^{\{w,x,y,z\}}) = (2, 2, \ldots, 2, 3, 3, 3, 3)$. It is easy to verify that this outcome is not blocked by any coalition. So the core is nonempty.

Now suppose there is no exact cover by 3-sets. Suppose the core is nonempty, that is, it contains some outcome $u^A = (u^{C_1}, \ldots, u^{C_r})$ with each $(C_i, u^{C_i})$ an element of $W$, and the $C_i$ disjoint. Then one of the $C_i$ must be $\{s, w\}$ for some $s \in S$: for if this were not the case, there must be some $s \in S$ with $u_s^A = 0$, because the $C_i$ that are equal to $S_i$ cannot cover $S$; but then $\{s, w\}$ would block the outcome. Thus, none of the $C_i$ can be equal to $\{w, x, y, z\}$. Then one of the $C_i$ must be one of $\{x, y\}, \{y, z\}, \{x, z\}$, or else two of $\{x, y, z\}$ would block the outcome. By symmetry, we can without loss of generality assume it is $\{x, y\}$. But then $\{y, z\}$ will block the outcome. (Contradiction.) So the core is empty. ∎

We might hope that the convexity introduced by transferable utility makes the problem tractable through, for example, linear programming. This turns out not to be the case.

**Theorem 2** *CORE-NONEMPTY with transferable utility is $\mathcal{NP}$-complete.*

**Proof**: To show that the problem is in $\mathcal{NP}$, nondeterministically choose a subset of $W$, and check if the corresponding coalitions constitute a partition of $A$. If so, nondeterministically divide the sum of the coalitions' values over the players, and check if this outcome is blocked by any element of $W$.

To show $\mathcal{NP}$-hardness, we reduce an arbitrary EXACT-COVER-BY-3-SETS instance to the following CORE-NONEMPTY instance. Let the set of players be $A = S \cup \{x, y\}$. For each $S_i$, let $(S_i, 3)$ be an element of $W$. Additionally, let $(S \cup \{x\}, 6m), (S \cup \{y\}, 6m)$, and $(\{x, y\}, 6m)$ be elements of $W$. The only other elements of $W$ are the required ones giving value 0 to singleton coalitions. We claim the two instances are equivalent.

First suppose there is an exact cover by 3-sets consisting of $S_{c_1}, \ldots, S_{c_m}$. Then the value of coalition $S$ is at least $\sum_{1 \leq i \leq m} v(S_{c_i}) = 3m$. Combining this with the coalition $\{x, y\}$, which has value $6m$, we conclude that the grand coalition $A$ has value at least $9m$. Hence, the outcome $(1, 1, \ldots, 1, 3m, 3m)$ is possible. It is easy to verify that this outcome is not blocked by any coalition. So the core is nonempty.

Now suppose there is no exact cover by 3-sets. Then the coalition $S$ has value less than $3m$ (since there are no $m$ disjoint $S_i$), and as a result the value of the grand coalition is less than $9m$. It follows that in any outcome, the total utility of at least one of $S \cup \{x\}, S \cup \{y\}$, and $\{x, y\}$ is less than $6m$. So this coalition will block. So the core is empty. ∎

Our results imply that it is computationally hard to make any strategic assessment of a game in characteristic form when it is concisely represented.

## 5 Specifying redundant information about the grand coalition makes the problem tractable

Our proofs that CORE-NONEMPTY is hard relied on constructing instances where it is difficult to determine what the grand coalition can accomplish. So, in effect, the hardness derived from the fact that even collaborative optimization is hard in these instances. While this is indeed a real difficulty that occurs in the analysis of characteristic form games, we may nevertheless wonder to what extent computational complexity issues are introduced by the purely strategic aspect of the games. To analyze this, we investigate the computational complexity of CORE-NONEMPTY when $V(A)$ (or $v(A)$) is explicitly provided as (possibly redundant) input, so that determining what the grand coalition can accomplish can no longer be the source of any complexity.[4] It indeed turns out that the problem becomes easy both with and without transferable utility.

**Theorem 3** *When $V(A)$ is explicitly provided, CORE-NONEMPTY without transferable utility is in $\mathcal{P}$.*

**Proof**: The following simple algorithm accomplishes this efficiently. For each element of $V(A)$, check whether it is blocked by any element of $W$. ∎

---

[4]Bilbao et al. have studied the complexity of the core in characteristic form games with transferable utility when there is an oracle that can provide the value $v(B)$ of any coalition $B$ [Bilbao *et al.*, ]. Our amended input corresponds to asking one such query in addition to obtaining the unamended input.

For the transferable utility case, we make use of linear programming.

**Theorem 4** *When $v(A)$ is explicitly provided, CORE-NONEMPTY with transferable utility is in $\mathcal{P}$.*

**Proof**: We decide how to allocate the $v(A)$ among the agents by solving a linear program. The core is nonempty if and only if the following linear program has a solution:
- $\sum_{1 \leq i \leq n} u_i \leq v(A)$;
- For any $(B, v(B))$ in $W$, $\sum_{b \in B} u_b \geq v(B)$. ∎

The algorithms in the proofs also construct a solution that is in the core, if the core is nonempty.

## 6 Hybrid games remain hard

Not all complexity issues disappear through having the collaborative optimization problem solution available. It turns out that if we allow for *hybrid* games, where only *some* coalitions can transfer utility among themselves, the hardness returns. In particular, we show hardness in the case where only the grand coalition can transfer utility. This is a natural model for example in settings where there is a market institution that enforces payments, but if a subset of the agents breaks off, the institution collapses so payments cannot be enforced.

We demonstrate $\mathcal{NP}$-hardness of this problem by reducing from the $\mathcal{NP}$-complete NODE-COVER problem [Garey and Johnson, 1979].

**Definition 10 (NODE-COVER)** *We are given a graph $G = (V, E)$, and a number $k$. We are asked whether there is a subset of $V$ of size $k$ such that each edge has at least one of its endpoints in the subset.*

We are now ready to state our result.

**Theorem 5** *When only the grand coalition can transfer utility, CORE-NONEMPTY is $\mathcal{NP}$-complete, even when $v(A)$ is explicitly provided as input.*

**Proof**: To show that the problem is in $\mathcal{NP}$, nondeterministically divide $v(A)$ over the players, and check if this outcome is blocked by any element of $W$.

To show $\mathcal{NP}$-hardness, we reduce an arbitrary NODE-COVER instance to the following CORE-NONEMPTY instance. Let $A = V \cup \{x, y, z\}$, and let $v(A) = 6|V| + k$. Furthermore, for each edge $(v_i, v_j)$, let $(\{v_i, v_j\}, u^{\{v_i, v_j\}})$ be an element of $W$, with $u^{\{v_i, v_j\}} = (1, 1)$. Finally, for any $a, b \in \{x, y, z\}$ ($a \neq b$), let $(\{a, b\}, u^{\{a,b\}})$ be an element of $W$, with $u^{\{a,b\}} = (3|V|, 2|V|)$. The only other elements of $W$ are the required ones giving utility 0 to singleton coalitions. This game does not violate the superadditivity assumption, since without the explicit specification of $v(A)$, superadditivity can at most imply that $v(A) = 6|V| \leq 6|V| + k$. We claim the two instances are equivalent.

First suppose there is a node cover of size $k$. Consider the following outcome: all the vertices in the node cover receive utility 1, all the other vertices receive utility 0, and each of $x$, $y$, and $z$ receives utility $2|V|$. Using the fact that all the edges are covered, it is easy to verify that this outcome is not blocked by any coalition. So the core is nonempty.

Now suppose there is some outcome $u^A$ in the core. In such an outcome, either each of $\{x, y, z\}$ receives at least $2|V|$, or two of them receive at least $3|V|$ each. (For if not, there is some $a \in \{x, y, z\}$ with $u_a^A < 2|V|$ and some $b \in \{x, y, z\}$ ($b \neq a$) with $u_b^A < 3|V|$, and the coalition $\{b, a\}$ will block through $u^{\{b,a\}} = (3|V|, 2|V|)$.) It follows that the combined utility of all the elements of $V$ is at most $k$. Now, for each edge $(v_i, v_j)$, at least one of its vertices must receive utility at least 1, or this edge would block. So the vertices that receive at least 1 cover the edges. But because the combined utility of all the elements of $V$ is at most $k$, there can be at most $k$ such vertices. So there is a node cover. ∎

Hybrid games, where only some coalitions can transfer utility, are quite likely to appear in real-world multiagent settings, for example because only some of the agents use a currency. Our result shows that for such hybrid games, even when the collaborative optimization problem has already been solved, it can be computationally hard to strategically assess the game.

## 7 Conclusions and future research

Coalition formation is a key problem in automated negotiation among self-interested agents, and other multiagent applications. A coalition of agents can sometimes accomplish things that the individual agents cannot, or can do things more efficiently. However, motivating the agents to abide to a solution requires careful analysis: only some of the solutions are stable in the sense that no group of agents is motivated to break off and form a new coalition. This constraint has been studied extensively in cooperative game theory. However, the computational questions around this constraint have received less attention. When it comes to coalition formation among software agents (that represent real-world parties), these questions become increasingly explicit.

In this paper we defined a concise general representation for games in characteristic form that relies on superadditivity, and showed that it allows for efficient checking of whether a given outcome is in the core. We then showed that determining whether the core is nonempty is $\mathcal{NP}$-complete both with and without transferable utility. We demonstrated that what makes the problem hard in both cases is determining the collaborative possibilities (the set of outcomes possible for the grand coalition), by showing that if these are given, the problem becomes tractable in both cases. However, we then demonstrated that for a hybrid version of the problem, where utility transfer is possible only within the grand coalition, the problem remains $\mathcal{NP}$-complete even when the collaborative possibilities are given.

Future research can take a number of different directions. One such direction is to investigate the complexity of restricted families of games in characteristic form.[5] Another

---

[5] One interesting restricted family is that of *convex games*. In a convex game (with transferable utility), for any $B, C \subseteq A$, $v(B) + v(C) \leq v(B \cup C) + v(B \cap C)$, and in such games the core is known to always be nonempty. How complex it is here to *construct*

direction is to evaluate other solution concepts in cooperative game theory from the perspective of computational complexity under our input representation. A long-term goal is to extend our framework for finding a strategically stable solution to take into account issues of computational complexity in determining the synergies among coalitions (for example, when routing problems need to be solved, potentially only approximately, in order to determine the synergies).

## References

[Aumann, 1959] R Aumann. Acceptable points in general cooperative n-person games. volume IV of *Contributions to the Theory of Games*. Princeton University Press, 1959.

[Bernheim *et al.*, 1987] B Douglas Bernheim, Bezalel Peleg, and Michael D Whinston. Coalition-proof Nash equilibria: I concepts. *J. of Economic Theory*, 42(1):1–12, June 1987.

[Bilbao *et al.*, ] J. M. Bilbao, J. R. Fernandez, and J. J. Lopez. Complexity in cooperative game theory.

[Bilbao, 2000] J. M. Bilbao. *Cooperative Games on Combinatorial Structures*. Kluwer Academic Publishers, 2000.

[Charnes and Kortanek, 1966] A Charnes and K O Kortanek. On balanced sets, cores, and linear programming. Technical Report 12, Cornell Univ., Dept. of Industrial Eng. and Operations Res., Ithaca, NY, 1966.

[Chatterjee *et al.*, 1993] K. Chatterjee, B. Dutta, D. Ray, and K. Sengupta. A noncooperative theory of coalitional bargaining. *Review of Economic Studies*, 60:463–477, 1993.

[Deng and Papadimitriou, 1994] Xiaotie Deng and Christos H. Papadimitriou. On the complexity of cooperative solution concepts. *Mathematics of Operations Research*, pages 257–266, 1994.

[Deng *et al.*, 1997] Xiaotie Deng, Toshihide Ibaraki, and Hiroshi Nagamochi. Algorithms and complexity in combinatorial optimization games. SODA 1997.

[Evans, 1997] Robert Evans. Coalitional bargaining with competition to make offers. *Games and Economic Behavior*, 19:211–220, 1997.

[Faigle *et al.*, 1994] U. Faigle, S. Fekete, W. Hochstattler, and W. Kern. On the complexity of testing membership in the core of min-cost spanning trees. Technical Report 94.166, Universitat zu Koln, 1994.

[Garey and Johnson, 1979] Michael R Garey and David S Johnson. *Computers and Intractability*. W. H. Freeman and Company, 1979.

[Gillies, 1953] D Gillies. *Some theorems on n-person games*. PhD thesis, Princeton, Dept. of Mathematics, 1953.

[Kahan and Rapoport, 1984] James P Kahan and Amnon Rapoport. *Theories of Coalition Formation*. Lawrence Erlbaum Associates Publishers, 1984.

[Ketchpel, 1994] Steven Ketchpel. Forming coalitions in the face of uncertain rewards. AAAI, pages 414–419, Seattle, WA, July 1994.

[Larson and Sandholm, 2000] Kate Larson, Tuomas Sandholm. Anytime coalition structure generation: An average case study. *J. of Experimental and Theoretical AI*, 11:1–20, 2000. Early version: AGENTS, pp. 40–47, 1999.

[Mas-Colell *et al.*, 1995] Andreu Mas-Colell, Michael Whinston, and Jerry R. Green. *Microeconomic Theory*. Oxford University Press, 1995.

[Milgrom and Roberts, 1996] Paul Milgrom and John Roberts. Coalition-proofness and correlation with arbitrary communication possibilities. *Games and Economic Behavior*, 17:113–128, 1996.

[Moreno and Wooders, 1996] Diego Moreno and John Wooders. Coalition-proof equilibrium. *Games and Economic Behavior*, 17:80–112, 1996.

[Okada, 1996] Akira Okada. A noncooperative coalitional bargaining game with random proposers. *Games and Economic Behavior*, 16:97–108, 1996.

[Ray, 1996] Indrajit Ray. Coalition-proof correlated equilibrium: A definition. *Games and Economic Behavior*, 17:56–79, 1996.

[Sandholm and Lesser, 1997] Tuomas Sandholm and Victor R Lesser. Coalitions among computationally bounded agents. *Artificial Intelligence*, 94(1):99–137, 1997. Special issue on Economic Principles of Multiagent Systems. Early version: International Joint Conference on Artificial Intelligence (IJCAI), pages 662–669, 1995.

[Sandholm *et al.*, 1999] Tuomas Sandholm, Kate Larson, Martin Andersson, Onn Shehory, and Fernando Tohmé. Coalition structure generation with worst case guarantees. *Artificial Intelligence*, 111(1–2):209–238, 1999. Early version appeared at the National Conference on Artificial Intelligence (AAAI), pages 46–53, 1998.

[Shapley, 1967] L. S. Shapley. On balanced sets and cores. *Naval Research Logistics Quarterly*, 14:453–460, 1967.

[Shehory and Kraus, 1996] Onn Shehory and Sarit Kraus. A kernel-oriented model for coalition-formation in general environments: Implemetation and results. In *Proceedings of the National Conference on Artificial Intelligence (AAAI)*, pages 134–140, Portland, OR, August 1996.

[Shehory and Kraus, 1998] Onn Shehory and Sarit Kraus. Methods for task allocation via agent coalition formation. *Artificial Intelligence*, 101(1–2):165–200, 1998.

[Tohmé and Sandholm, 1999] Fernando Tohmé and Tuomas Sandholm. Coalition formation processes with belief revision among bounded rational self-interested agents. *Journal of Logic and Computation*, 9(97–63):1–23, 1999. Early version: IJCAI-97 Workshop on Social Interaction and Communityware, pp. 43–51, Nagoya, Japan.

[van der Linden and Verbeek, 1985] Wim J van der Linden and Albert Verbeek. Coalition formation: A game-theoretic approach. In Henk A M Wilke, editor, *Coalition Formation*, volume 24 of *Advances in Psychology*. North Holland, 1985.

---

a solution in the core with our representation is an open question.

[Wu, 1977] L. S. Wu. A dynamic theory for the class of games with nonempty cores. *SIAM Journal of Applied Mathematics*, 32:328–338, 1977.

[Zlotkin and Rosenschein, 1994] Gilad Zlotkin and Jeffrey S Rosenschein. Coalition, cryptography and stability: Mechanisms for coalition formation in task oriented domains. In *Proceedings of the National Conference on Artificial Intelligence (AAAI)*, pages 432–437, Seattle, WA, July 1994.